\documentclass{article}

\usepackage{amssymb}
\usepackage{bbm}
\usepackage{color}
\usepackage{colortbl}
\usepackage{graphicx}
\usepackage{multirow}
\usepackage{natbib}
\usepackage{vmargin}

\RequirePackage[OT1]{fontenc}
\RequirePackage[colorlinks,citecolor=blue,urlcolor=blue,bookmarks=false]{hyperref}

\begin{document}

\title{A penalized likelihood approach to estimate within-household contact networks from egocentric data}
\author{Gail E. Potter and Niel Hens}
\date{}
\maketitle

\begin{abstract}
Acute infectious diseases are transmitted over networks of social contacts.  Epidemic models are used to predict the spread of emergent pathogens and compare intervention strategies.  Many of these models assume equal probability of contact within mixing groups (homes, schools, etc.), but little work has inferred the actual contact network, which may influence epidemic estimates.  We develop a penalized likelihood method to infer contact networks within households, a key area for disease transmission.  Using egocentric surveys of contact behavior in Belgium, we estimate within-household contact networks for six different age compositions.  Our estimates show dependency in contact behavior and vary substantively by age composition, with fewer contacts occurring in older households.  Our results are relevant for epidemic models used to make policy recommendations.  
\end{abstract}

\section{Introduction}

Acute infectious diseases, such as influenza, spread through networks of face-to-face social contacts.  When a new strain of influenza virus emerges, a variety of epidemic models are used to estimate key epidemic parameters, simulate and predict epidemic spread, and compare intervention strategies.  The majority of these models are based on the simplistic ``random mixing'' assumption regarding social contact behavior.  Under this assumption, people contact each other with equal probability within mixing groups (homes, schools, workplaces, etc.), but no other social contact structure is modeled.  For example, the large scale agent-based models in~\cite{eubank1},~\cite{germann},~\cite{Ferguson06n}, and~\cite{halloran} assume random mixing within homes, grades and/or schools, workplaces and workgroups, and communities.  Furthermore, random mixing within households is used in models estimating secondary attack rates within households.  See~\cite{Longini88aje, halloran2007, Yang07csda} and~\cite{yang}.   Classical models to estimate the basic reproductive number $R_0$ assume random mixing with age-specific contact probabilities (e.g.~\cite{diekmann} and~\cite{anderson}).  Because these models use infection or symptom data but not contact data, age differentials in their transmission rate estimates result both from differential infectiousness and susceptibility by age, as well as differences in contact behavior by age.  An understanding of the contact network is essential to disentangle the effects of biology and behavior.

Researchers have demonstrated that network structure can result in different epidemic predictions than random mixing.~\cite{Keeling05jrsi} reviewed idealized types of networks which have been used to approximate the contact network, and compared the epidemic curve from simulated disease transmission over various network types to that obtained over random mixing.~\cite{Keeling05jrsi} and~\cite{Miller09jrsi} showed that clustering affects the course of the epidemic and explored how the effect varies by clustering level and for different types of networks.  Researchers are actively involved in estimating properties of contact networks and integrating survey-based network information into epidemic estimation models.~\cite{wallinga} supplemented infectious disease data with social contact data to improve estimates of age-specific transmission parameters.  They demonstrated that their model, which integrates the age-specific contact rates and mixing patterns, improves model fit over random mixing.~\cite{goeyvaerts} extended the methodology in~\cite{wallinga} and applied it to the Belgian data from the POLYMOD study, a multi-country European survey of contact behavior, which we analyze in this paper.  In addition,~\cite{goeyvaerts2011} combined social contact data with serological data for human parovirus B19 (PVB19) and found evidence for age-specific waning of PVB19 immunity in four of five European countries they analyzed.

Households are known to be a primary component of the disease transmission process, but relatively little work has been done to estimate contact networks within households.  As mentioned previously, most household models assume random mixing within households.~\cite{britton} developed a Bayesian method to estimate the infection rate, mean of the infection period, and probability of social contact, and assumed this probability is equal for all pairs; i.e. random mixing.~\cite{demiris} developed inference for infection rates and imputed the contact graph, assuming random mixing within and between groups.~\cite{potter} is the first paper we know of that develops inference for within-household contact networks from egocentric data.  
They applied their parametric model to the same data set we analyze here.  

We contribute to this area by developing a method to estimate small contact networks from survey data and applying it to model networks of household contacts using the Belgian POLYMOD data.  We estimate the probability distributions for household networks of size four of various age compositions in Belgium.  We compare the results to a random mixing scenario, and we investigate the effect of age composition on the contact network.  Our method requires fewer assumptions about contact behavior than that of~\cite{potter}.

Our method also contributes to the field of social network methodology by inferring the probability distribution for complete networks from partially observed network data.  We represent a network graphically by using nodes to represent social actors and ties to represent contacts between people, and mathematically by a square matrix $\mathbf{Y}$ where $Y_{ij}=1$ if persons $i$ and $j$ make contact and $Y_{ij}=0$ if not.  One standard class of network models, exponential family random graph models (ERGMs) represent global network structure as a function of local social behavior~\citep{strauss}.  Inference for ERGMs was developed assuming observation of the complete network;~\cite{markandkrista} developed inference for ERGMs from partially observed networks.  Such estimation assumes that the ERGM is correctly specified: that the features of the network are indeed captured by the network statistics included in the model.  For exploratory work to describe an unknown network or get an initial sense of which statistics will be relevant, a nonparametric estimation procedure of the probability distribution would be very useful.

The network data we analyze is egocentric: randomly sampled respondents were interviewed about their contacts to other members, but they did not report on contacts between other members.  They reported attributes of people they contacted but not identities.  Egocentric data is a commonly available network data type.  It contains information about assortative mixing (the tendency to contact others with similar attributes) and the degree distribution, where the degree is the number of contacts a person makes.  Egocentric data does not include information about transitivity or other higher-level network structures.  Network inference for egocentric data may be performed by assuming contacts occur independently conditional on individual-level attributes (as described in~\cite{koehly}), or by imposing a dependence structure.  We ascertain the identities of household contacts by matching the age of the contacted member to the household age roster.  Thus, our data set contains more information than a random egocentric sample, permitting us to estimate dependence in contact behavior.  The networks we analyze are size four with a single respondent per household.  Thus, each respondent reported half of the network (three of six possible contacts).  Reports from different respondents in multiple households therefore contain a fair amount of information to characterize the probability distribution of the network.  

This paper is organized as follows.  In section~\ref{section:data}, we describe the POLYMOD study.  In section~\ref{section:nonparametric} we present a nonparametric maximum likelihood method to estimate the probability distribution of a small contact network of fixed size from egocentric data.  With the constraint of assuming that children are exchangeable and adults are exchangeable, this method can be used to estimate the nonparametric MLE of the contact network distribution for a large data set, but in smaller data sets such as ours, the parameters are not identifiable.  We resolve this through a penalized likelihood approach, described in section~\ref{section:penalized}.  Our penalty imposes a mathematical preference for distributions representing networks where contacts between members occur independently of each other.  In section~\ref{section:simulation} we describe a simulation study to assess predictive performance of our method in large data sets.  We estimate the probability distribution of within-household contact networks for households of size four of six different age compositions in Belgium.  Estimates for three household types are presented and compared in section~\ref{section:estimates}; we also compare the estimates to random mixing.  Results from the three other household compositions are in the supplementary material.  Results from the simulation study are presented in section~\ref{section:sim results}.  In section~\ref{section:discussion} we discuss our findings and the performance of our method.

\section{The POLYMOD Data}
\label{section:data}

The POLYMOD survey was administered in eight European countries in 2006 and contains detailed diaries of contact behavior during a day.  We analyze the Belgian POLYMOD data. ~\cite{mossong} analyzed the POLYMOD data set and compare contact patterns between countries, and~\cite{hens2009} analyzed the Belgian POLYMOD data using association rules and classification trees. In Belgium, random digit dialing was used to obtain consent, and sampling weights ensure that the three main regions of Belgium were represented (Flemish, Walloon, and Brussels).  Children were oversampled because they are key transmitters of infections.  Data were collected from 750 respondents during March--May of 2006, with one respondent per household.  Each respondent was mailed a paper diary and was assigned two randomly selected days, one weekday and one weekend day.  To ensure that observations are independent, we analyze the first day reported by each respondent.  Approximately half of respondents (381 of 750) filled out the first day of their diary during the two-week Easter holiday  period (April 3--17), during which schools were closed.  For each assigned day, respondents were instructed to record information about all social contacts from 5 a.m. till 5 a.m. the next morning.  A contact was defined to be a two-way conversation of at least three words in the same location and/or a physical contact.  The diary includes one row for each person contacted on a given day.  The age and sex of the person contacted were recorded, as well as attributes of the contact itself including frequency (daily or almost daily, once or twice a week, once or twice a month, less than once a month, or for the first time), and location (home, work, school, leisure, transport, or other).  Respondents also listed demographic information of self and their households, including ages of all household members.  

Respondents did not report whether people contacted were household members or not, and our aim is to estimate the contact networks between household members.  We assume that contacts were to household members if they occurred ``at home'', were reported as ``daily or almost daily'', and if their age matches one of the reported ages of household members.  For each household we observe a partial contact network: we have information on ties between the respondent and all other members, but not on contacts between other members.  Our data is egocentric, but with the assumptions we have made, includes the identity of the alters.

We develop a method to model the contact network for households of fixed size and age composition and apply this method to households of size four in the Belgian POLYMOD data.  We classify members into the following age categories which we expect to exhibit different contact behavior:  0--5, 6--11, 12--18, 19--35, and 36+.  Table~\ref{tab:composition} shows the distribution of age compositions of households of size four in our data set.  



%

Table~\ref{tab:composition2} shows the six household composition types we analyze in this paper.  In households with small children, we collapsed the two adult age groups to obtain adequate sample sizes for each group.  Based on our understanding of social norms, we expect each of these households to exhibit different contact patterns.

\begin{table}
\caption{\label{tab:composition} Age composition of households of size four in the Belgian POLYMOD data set.}
\centering
\begin{tabular}{cccccc}
\hline 
\multicolumn{5}{c}{Age Category}& \multicolumn{1}{c}{Number of}\\
0--5 & 	6--11 & 	12--18 & 19--35 & 36+ & Respondents\\ 
\hline 
0 & 	0 & 	0 & 	0 & 	4 & 	1 \\ 
0 & 	0 & 	0 & 	1 & 	3 & 	1 \\ 
0 & 	0 & 	0 & 	2 & 	2 & 	35 \\ 
0 & 	0 & 	0 & 	3 & 	1 & 	1 \\ 
0 & 	0 & 	0 & 	4 & 	0 & 	1 \\ 
0 & 	0 & 	1 & 	1 & 	2 & 	23 \\ 
0 & 	0 & 	1 & 	2 & 	1 & 	1 \\ 
0 & 	0 & 	2 & 	0 & 	2 & 	40 \\ 
0 & 	0 & 	3 & 	0 & 	1 & 	2 \\ 
0 & 	1 & 	0 & 	0 & 	3 & 	1 \\ 
0 & 	1 & 	0 & 	1 & 	2 & 	1 \\ 
0 & 	1 & 	1 & 	1 & 	1 & 	2 \\ 
0 & 	1 & 	2 & 	0 & 	1 & 	1 \\ 
0 & 	1 & 	1 & 	0 & 	2 & 	17 \\ 
0 & 	1 & 	2 & 	0 & 	1 & 	1 \\ 
0 & 	2 & 	0 & 	0 & 	2 & 	16 \\ 
0 & 	2 & 	0 & 	1 & 	1 & 	8 \\ 
0 & 	2 & 	0 & 	2 & 	0 & 	4 \\ 
1 & 	0 & 	1 & 	0 & 	2 & 	1 \\ 
1 & 	1 & 	0 & 	0 & 	2 & 	6 \\ 
1 & 	1 & 	0 & 	1 & 	1 & 	8 \\ 
1 & 	1 & 	0 & 	2 & 	0 & 	12 \\ 
2 & 	0 & 	0 & 	0 & 	2 & 	2 \\ 
2 & 	0 & 	0 & 	1 & 	1 & 	12 \\ 
2 & 	0 & 	0 & 	2 & 	0 & 	16 \\ 
\hline 
\end{tabular}
\end{table}
\begin{table}
\caption{ \label{tab:composition2} Household composition types analyzed in this paper}
\centering
\begin{tabular}{cccccc}
\hline 
Household Type &Child 1 &Child 2 &Parent 1 &Parent 2& n\\
\hline
Type 1& 0-5   &0-5   &19+ &19+ &30\\
Type 2& 0-5   &6-11 &19+  &19+ &26\\
Type 3& 6-11  &6-11 & 19+ &19+ &28\\
Type 4& 12-18 &12-18 &36+ &36+ &40\\
Type 5& 12-18 &19-35 &36+& 36+ &23\\
Type 6& 19-35 &19-35 &36+& 36+ &35\\
\hline 
\end{tabular}
\end{table}

\begin{figure}[tbp] 
  \caption{Subset of observed data: households with two 0--5 year olds and two 19+ year olds; respondent in blue.  Lines indicate reported contact.  Labels are: ch=child, ad=adult. \label{fig:data}}  
\centering
  \includegraphics[width=3.94in,height=2.65in,keepaspectratio]{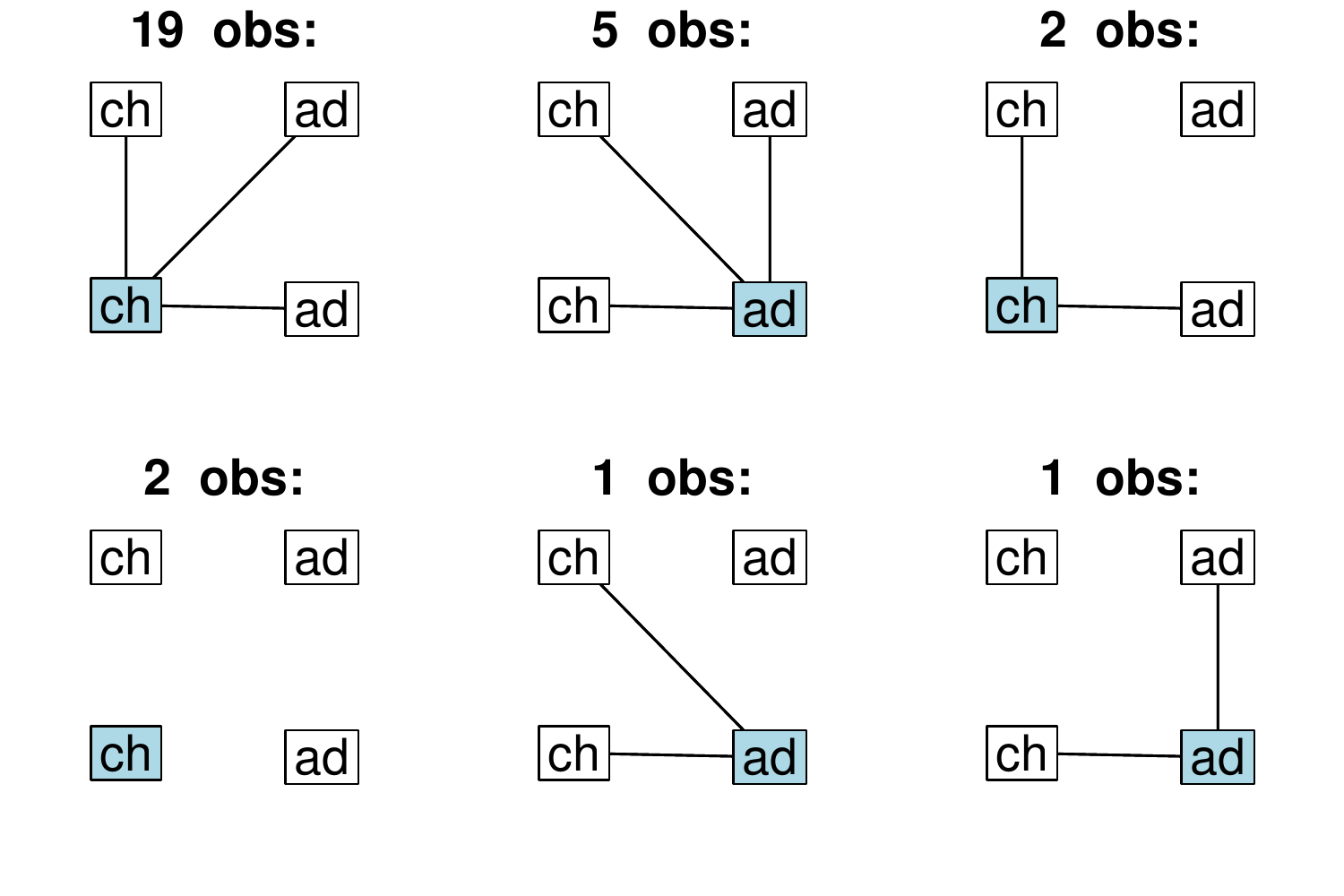}
\end{figure}

\begin{figure}[tbp] 
  \caption{Subset of observed data: households with two 19--35 year olds and two 36+ year olds; respondent in blue.  Lines indicate reported contact.  Labels are: ch=child, ad=adult.} 
  \centering
  \includegraphics[width=5.43in,height=6.07in,keepaspectratio]{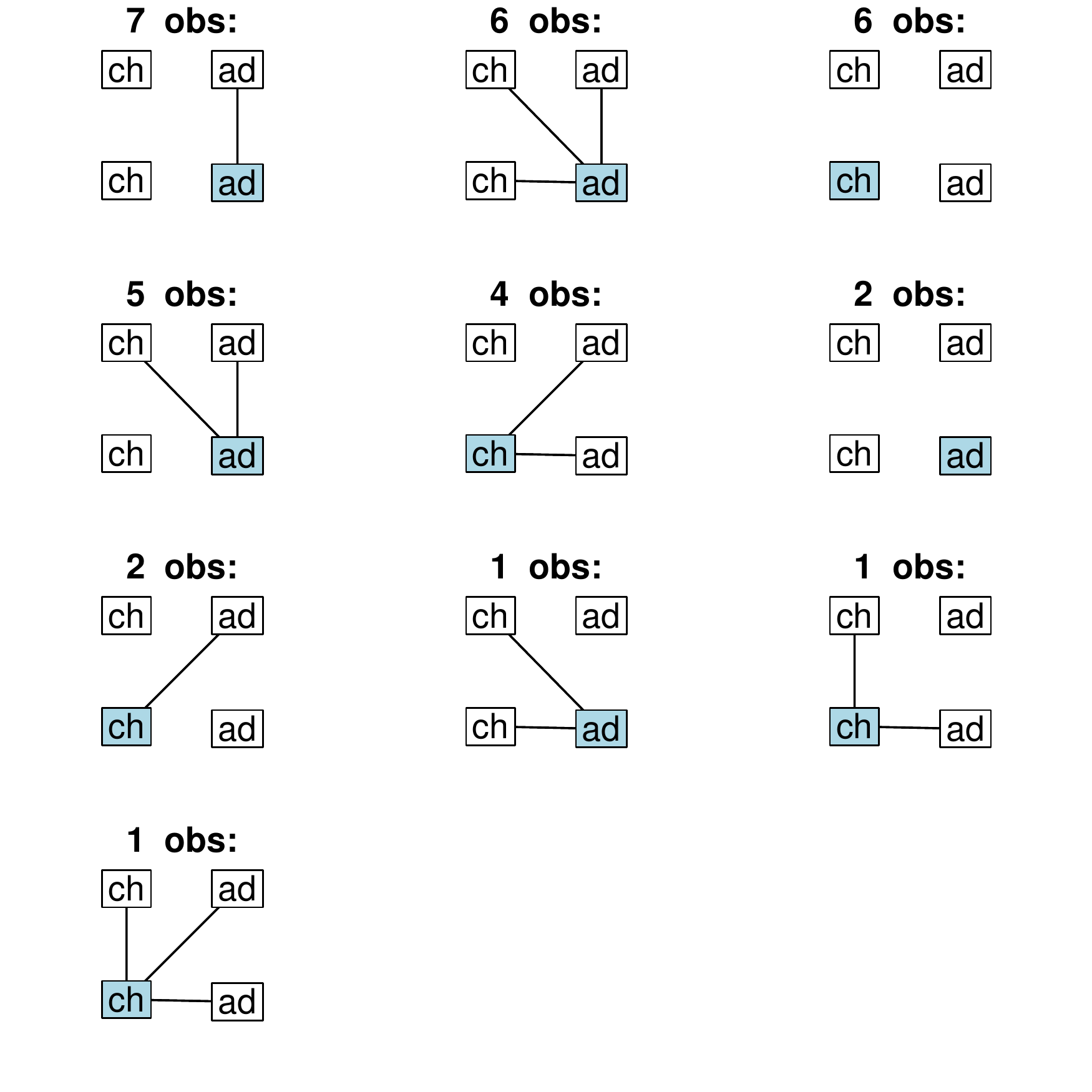}
\label{fig:data2}
\end{figure}

Figure~\ref{fig:data} shows our observed data for households with two 0--5 year olds and two 19+ year olds.  The respondent is marked in blue, and ties indicate reported contacts.  Because of our structurally missing data, contact status on dyads excluding the respondent is not observed.  In order to display the observed data concisely, we assume that the two children are exchangeable and the two adults are exchangeable.  However, we do not make this assumption in our model.  Our model allows the younger child to behave differently from the older child and the female adult to behave differently from the male adult.  Figure~\ref{fig:data2} shows observed data for households with two young adults and two older adults.  Density of contact is substantially smaller than it is in the younger household type, and we see more diverse reporting patterns in this type of household. 

\section{Methodology}
\subsection{A Nonparametric Approach}
\label{section:nonparametric}

We develop a technique for estimating the probability distribution of a small household network of fixed size from egocentric data.  The method makes no assumptions about the similarity in behavior between household members.  Here we discuss its application to a household of size four with two 0--5 year olds, and two 19+ year olds.  
Contacts as defined by the survey are symmetric, so there are ${4\choose 2}=6$ possible contacts in each household.  We will use vector notation to represent the network, since it is more compact than matrix notation and easier to display our results.  We represent the household network by a 6-vector, $z$, where each element of $z$ represents a possible contact between two members.  The total number of possible contact networks for a household of this age composition is thus $2^6 = 64$.  

For each surveyed household, only three of the six possible contacts are observed, since respondents reported on their contacts to other household members but not on contacts between other household members.  Let $y$ denote the observed network, a 6-vector where three elements are missing.  

We first express the likelihood of the data in the most general form, which allows for any parametrization.  Let $Y_i$ denote the vector representing the network reported by respondent $i$, and let $n$ be the number of respondents.  Let $R_i$ denote the respondent type of respondent $i$, which takes one of the following four values: younger child, older child, female adult, male adult.  We denote the probability of network $k$ by $p_\theta(k)$, where $\theta$ is a vector of parameters.  Sampling probabilities of the various respondent types are denoted $p_\psi(R_i=r_i)$.  The separate parametrization of the network probability distribution and the sampling probabilities is justified by the sampling design: the process of selecting respondents was independent of the within-household contact network. 

Each observation includes the respondent type which determines which dyads are observed, as well as the values of the observed dyads.  We can compute the likelihood contribution of one respondent by summing the probabilities of all complete networks which are consistent with the partially observed network.   The joint probability mass function of observed respondent type and observed dyadic data is thus

\[
P(Y_i = y_i, R_i=r_i | \theta, \psi ) = \Bigg( \sum_{k=1}^{64} p_\theta (k) 1_\textrm{[k,i]} \Bigg) p_\psi(R_i = r_i),
\]
where 
\[
1_\textrm{[k,i]} = \left\{
\begin{array}{rl}
1 & \textnormal{if partially observed network $y_i$ is consistent with network k}\\
0 & \textnormal{otherwise}
\end{array} \right.
\]

The joint likelihood function of $\theta$ and $\psi$ is thus:
\[
L(\theta, \psi | Y_i = y_i, R_i=r_i) = \Bigg( \sum_{k=1}^{64} p_\theta (k) 1_\textrm{[k,i]} \Bigg) p_\psi(R_i = r_i)
\]

We are concerned with estimation of $\theta$, and it's clear that the score equations for $\theta$ will be free of 
$\psi$.  Thus we can restrict our attention to the likelihood for $\theta$ alone:

\[
L(\theta | Y_i = y_i, R_i=r_i) \propto \sum_{k=1}^{64} p_\theta (k) 1_\textrm{[k,i]} 
\]

We begin by describing a nonparametric approach, in which we assume no functional relationship between the probabilities of different networks; that is $p_\theta (k)\equiv p_k$, where $\mathbf{p}$ is a vector in 64-space.  This approach makes no assumptions about the similarity of contact behavior between household members.  The likelihood of $\mathbf{p}$ is thus 

\[
L(\mathbf{p} | Y_1 = y_1, \ldots Y_n=y_n,\mathbf{R}=\mathbf{r}  ) \propto \prod_{i=1}^{n} \sum_{k=1}^{64} 
p_k 1_\textrm{[k,i]} 
\]

We would like to maximize the above likelihood function to obtain the maximum likelihood estimate (MLE), but we have an identifiability problem.  
The likelihood function includes 63 free parameters (64 which sum to one).  The number of possible distinct data configurations is 32, as
there are four types of respondents (so four missingness patterns) and $2^3=8$ possible reports from each respondent.  Estimation will only
be possible if we can restrict our parameter space to have 32 or fewer free parameters.  
One way to reduce the identifiability problem is to assume that the two children are exchangeable and the two adults are exchangeable.  This reduces the dimension of the parameter space to 27 (28 parameters which sum to one).  However, we feel this approach 
is sensible only when the two children fall into the same age group, so the method could not be applied to households with two children in
different age groups.  In addition, we expect the female and male adults in the household to behave differently.  Moreover, we still do not have enough observed data points to accurately estimate the parameters.  Although there are 27 types of data configurations, only nine of these possibilities are observed in our data set with two 0--5 year olds and two 19+ year olds.  Our data does not contain enough information to estimate all the parameters in the likelihood.  

\subsection{A Penalized Likelihood Approach}
\label{section:penalized}

To resolve the identifiability problem, we use a penalized likelihood approach, also referred to as regularization~\citep{kim}.  We add to the likelihood a smoothing penalty which imposes a preference for probability distributions of networks in which contacts occur independently, a common assumption in epidemic models.  

When we assume independence, we have only six parameters, the probabilities of contact between each pair of household members.  We'll denote them by $\eta$, a vector with six elements.  We estimate $\eta_j$ with the MLE of the binomial distribution: 
\[
\hat{\eta}_j = \frac{\sum_{i=1}^n 1_{[d_{j,i}=1]} }{\sum_{i=1}^n 1_{[d_{j,i}=0]}+\sum_{i=1}^n 1_{[d_{j,i}=1]}},
\]
where $d_{j,i} =1$ if respondent $i$ reports contact on dyad $j$, $d_{j,i} =0$ if non-contact is reported, and $d_{j,i}$ is not observed for all respondents due to the structurally missing data.  

When we assume independence, the probabilities of each network are a deterministic function of $\eta$:
\[
\textnormal{P(}\mathbf{Z}=\mathbf{z})=\prod_{j=1}^6 \eta_j^{z_j} (1-\eta_j)^{1-z_j}
\]

Let $p_{k,ind}$ denote the probability of network $k$ under the independence assumption as described above, while as mentioned previously, $p_k$ denotes the unknown probability of network $k$ with no independence restriction.  We use the squared Hellinger distance to compare these distributions, so our penalized likelihood function with the independence penalty is:

\[
PL(\mathbf{p},\lambda) = \textrm{log}L(\mathbf{p}|y_1,\ldots y_n)) - 
\lambda   \Big( \frac{1}{2} \sum_{k=1}^{64} (\sqrt{p_{k,ind}} - \sqrt{p_k} )^2 \Big),
\]

The tuning parameter, $\lambda$, controls the degree of smoothness that is applied to the likelihood.  When $\lambda=0$, the estimates are completely informed by the data without any parametric assumptions.  As $\lambda\to\infty$, the penalty dominates the formula, and our estimate converges to the independence estimate. 

The choice of penalty may influence the results.  We tried two other penalty functions and compared their effect on the results.  We tried a penalty which imposes a preference for distributions in which networks differing on a single dyad have similar probabilities, defined by:
\[
PL(\mathbf{p},\lambda) = \textrm{log}L(\mathbf{p}|y_1,\ldots y_n)) - \lambda \sum_{i,j} (p_i - p_j )^2 1_{[\textrm{net i and net j differ on a single dyad}]}
\]

As expected, this penalty smooths the probability parameters, but we found the extent of smoothing to result in unrealistic estimates of probability distributions.  Results are included in the supplementary material.  

We also tried a penalty which imposes a preference for probability distributions in which the two children are exchangeable and the two adults are exchangeable.  We define the penalized log likelihood function with this penalty as follows:

\[
PL(\mathbf{p},\lambda) = \textrm{log}L(\mathbf{p}|y_1,\ldots y_n)) - 
\lambda \sum_{i,j} (p_i - p_j)^2 1_{[\textrm{net $i$, net $j$ isomorphic under exchangeability}]}
\]

We found that this penalty does not contribute enough information to resolve our identifiability problem.  There are a total of 28 unique networks when accounting for isomorphisms under exchangeability, but our subset of households with two 0--5 year olds and two 19+ year olds contains only nine types of partially observed networks.  Thus, even with a very large tuning parameter, the exchangeability penalty is insufficient to identify the parameters.  

To select the tuning parameter, we performed leave-one-out cross-validation (CV) as described by~\cite{hastie}. We implemented the procedure as follows:

We performed the following algorithm for $\lambda$ on a grid ranging from 0 to 40:
\begin{enumerate}
\item{}	Omit one data point, maximize the penalized likelihood for the remaining $n-1$.
\item{}	For the (penalized) MLE, compute the non-penalized likelihood for the omitted point.
\item{}	Repeat (1) and (2) $n$ times, so that each data point is omitted for one iteration.
\item{}	Compute the mean of the non-penalized likelihood over all $n$ iterations.
\end{enumerate}
We selected the value of $\lambda$ which maximized the mean of the non-penalized likelihood.  This is an extension of cross-validation from a prediction setting to a likelihood setting, in which we replace minimization of mean squared error with maximization of the likelihood.
 
An alternate way to define the optimal tuning parameter is the smallest $\lambda$ which results in an identifiable penalized likelihood.  According to~\cite{catchpole}, we can measure the identifiability of a likelihood equation by the rank of the Hessian matrix at the MLE, for exponential families.  We tried this approach as well, but found a large amount of noise in the relationship between the rank of the Hessian and the tuning parameter.  We expect this relationship to be monotone and positive.  Computing the rank of the true (rather than observed) 63 by 63 Hessian matrix is a non-trivial problem and is beyond the scope of this paper.  Thus, we present only results using the cross-validation-selected $\lambda$.  

We maximized the penalized likelihood function, subject to the constraint that the probabilities sum to 1 and all lie between 0 and 1, to obtain the penalized maximum likelihood estimate.  We performed optimization in R version 2.9.2~\citep{R}, with the {\tt optim} function and the BFGS method, discovered simultaneously by ~\cite{broyden},~\cite{fletcher},~\cite{goldfarb}, and~\cite{shanno}.  We compute standard errors for the penalized likelihood estimates by inverting the Fisher information matrix~\citep{lehcas}.  The Fisher information matrix is estimated with the {\tt optim} function in R when the {\tt hessian} argument to the function is set equal to true.  The asymptotic theorem requires identifiability of the parameter vector, and the estimated information matrix will not, in general, be invertible when the parameter vector is not identifiable.  An alternate method for estimating uncertainty with fewer assumptions is the nonparametric bootstrap, as described by~\cite{efron}.  We apply the nonparametric bootstrap to estimate standard errors for the unpenalized MLE, using 150 bootstrap resamples.  For comparison purposes, we also computed estimates and confidence intervals for the independence model described above.  Under the independence assumption, the number of contacts for each combination of household member types follow independent binomial distributions with parameters $\eta_1, \ldots, \eta_6$.  For some dyads, 100\% of contacts were observed, so Wald or score confidence intervals cannot be used.  The bootstrap would underestimate uncertainty for these parameters since each resample would also include 100\% of contacts observed.  Instead, for the independence model alone, we computed confidence intervals as follows: First, we computed exact binomial confidence intervals for each $\eta_j$, which we denote $[\eta_{j,low}, \eta_{j,up}]$.  We then computed a conservative 95\% confidence interval for each network $z$ as \[
\textnormal{CI low}=\prod_{j=1}^6 \eta_{j,low}^{z_{j}} (1-\eta_{j,up})^{1-z_j} \hspace{5mm} \textnormal{and} \hspace{5mm}
\textnormal{CI up}=\prod_{j=1}^6 \eta_{j,up}^{z_j} (1-\eta_{j,low})^{1-z_j}
\]

\subsection{Simulation Study}
\label{section:simulation}

We performed a simulation study to assess predictive performance of our method as follows.  We generated 200 samples of 30 networks from the unpenalized MLE for households with two 0--5 year olds and two 19+ year olds, since there are 30 households of that composition in our data set.  Next, we randomly assigned respondent status to one person in each simulated household using the observed frequency of different respondent types: six younger children, 17 elder children, four female adults, and three male adults.  We recoded dyads which would not be reported by the respondent as missing.  The penalized likelihood approach was then used to estimate the multinomial probability vector for a grid of $\lambda$-values ranging from 0 to 50 by steps of 0.5.  Based on the estimated probability vector we computed the mean average squared error and its bias-variance decomposition using the following definitions:

\begin{eqnarray*}
\mbox{MSE}(\lambda)&=&\frac{1}{64}\sum_{k=1}^{64} \frac{1}{200}\sum_{s=1}^{200} \left(\hat{p}_{sk}(\lambda)-p_{\mbox{true},k}\right)^2,\\
\mbox{Bias}(\lambda)&=&\frac{1}{64}\sum_{k=1}^{64} \left(\overline{\hat{p}}_{k}(\lambda)-p_{\mbox{true},k}\right),\\
\mbox{Variance}(\lambda)&=&\frac{1}{64}\sum_{k=1}^{64} \frac{1}{200}\sum_{s=1}^{200} \left(\hat{p}_{sk}(\lambda)-\overline{\hat{p}}_{k}(\lambda)\right)^2\\
\end{eqnarray*}

We repeated this procedure using the unpenalized MLE from a different household type: households with two 12--18 and two 36+ year olds, using the observed sample size (40) and respondent frequency (8 younger children, 20 elder children, 4 female adults, and 8 male adults) for this household composition.  We performed the simulation study with the independence penalty and the adjacency penalty.  For the adjacency penalty, we performed simulations for $\lambda$-values ranging from 0 to 10 by steps of 0.25, because the trends in bias, MSE, and variance are more visible in this range.  

\section{Results}
\label{section:results}

\subsection{Penalized likelihood estimates}
\label{section:estimates}

\begin{figure}[tbp] 
  \caption{Cross-validation results for the independence penalty\label{fig:gcvhell_squared}}
  \centering
 \includegraphics[width=3.24in,height=3.21in,keepaspectratio]{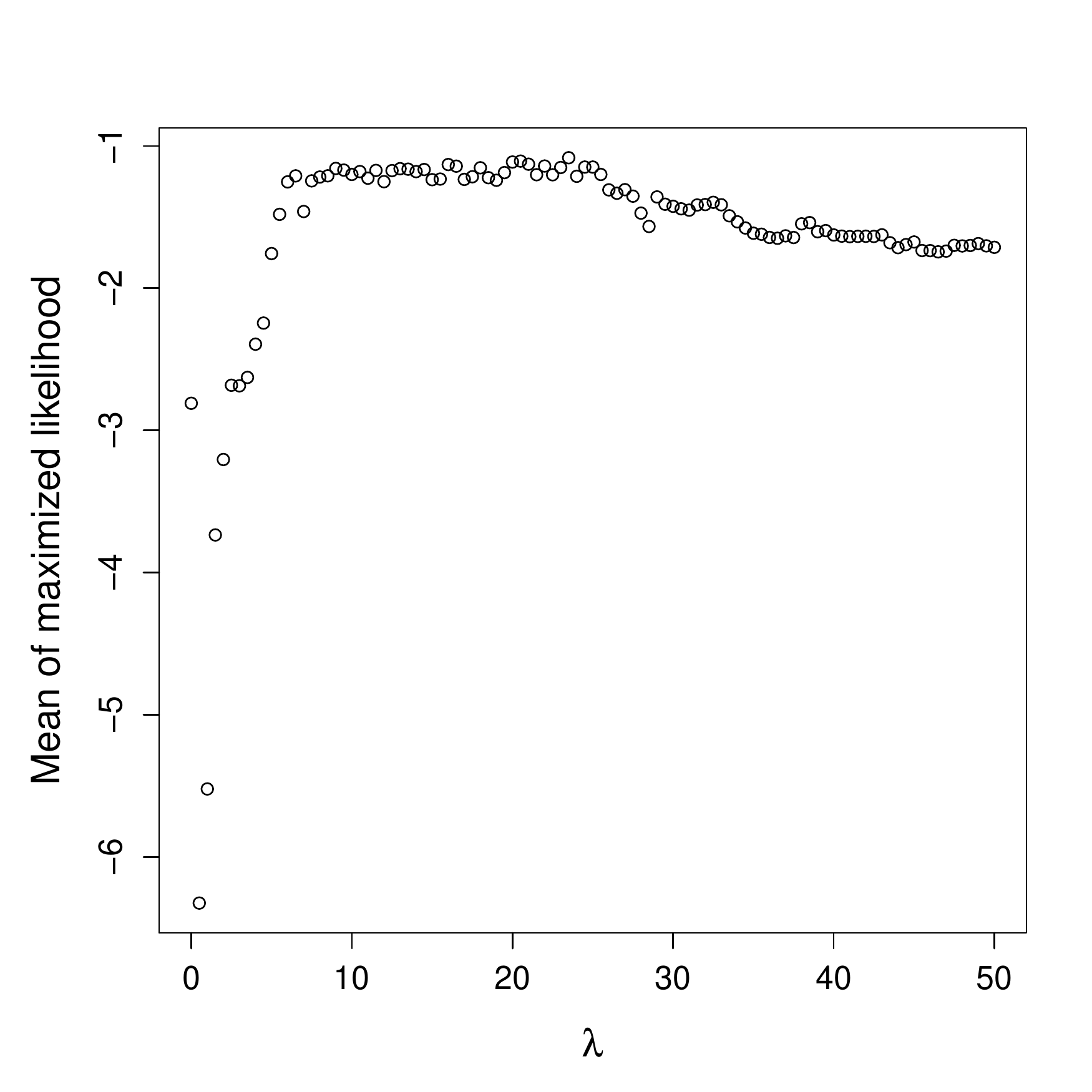} 
\end{figure}

Figure~\ref{fig:gcvhell_squared} shows the relationship between the tuning parameter and the mean of the likelihood from the cross-validation procedure for households with two 0--5 year olds and two 19+ year olds.  The maximum occurs at $\lambda=23.5$.  As expected, the curve is concave down, although there is more noise than expected.  Other household compositions showed less noise in the relationship; those plots are included in the  supplementary material.

Table~\ref{tab:hellindG} shows estimates for the probability distribution of the network estimated in three different ways: unpenalized MLE, independence MLE, and penalized MLE with CV-estimated $\lambda=23.5$.  To ease comparison of estimates between the three household types, we display the estimates in adjacent columns, followed by confidence intervals in adjacent columns.  We omit from display networks whose probability estimates under all three models were less than 0.02.  For this household composition the complete network (in which all contacts occur) receives a high probability estimate by all three methods.  As we would expect, the penalized likelihood estimates generally lie between the unpenalized estimates and the independence estimates.  The second network in the table receives non-negligible probability mass under both the penalized and unpenalized methods, but zero probability under the independence model.  This indicates that the data give support for this network, but the restrictions of the independence model are too strong to detect that support.  The smoothing imposed by the penalty does not remove the preference for this network.  

Our estimates show that the random mixing assumption is violated.  Random mixing assumes independence in contact behavior, as well as a single contact probability for all pairs.  Under random mixing, networks with the same number of total contacts have equal probability.  The independence model shows departure from random mixing, since networks with five total contacts have estimated probabilities 0.05, 0.13, 0.10, and 0.09 respectively.  Because the penalized MLE differs from the independence estimates, it shows evidence for dependence in contact behavior, indicating further departure from random mixing.  

Table~\ref{tab:hellindC} shows the estimates for households with two 12--18 year olds and two 36+ year olds.  Because the CV-selected $\lambda=199$ was much larger for this household, the penalized likelihood estimates are closer to the independence estimates.  However, they are still distinct.  The probability estimate for the complete network increases from 0.22 to 0.34.  Tables of estimates for the other four household composition types are included in the supplementary material.  The values of $\lambda$ estimated by cross-validation varied from 20 to 199.  Small values of $\lambda$ suggest that the data set contributes a fair amount of predictive power, so less smoothing is necessary.  Larger values of $\lambda$ show the need for more smoothing.  

\begin{table}
\caption{\label{tab:hellindG}Estimated probability distribution of contact network for households with two 0--5 year olds and two 19+ year olds. 
Dyad-independent, penalized likelihood (CV-selected $\lambda=23.5$), and unpenalized likelihood estimates are shown.}
\centering
\begin{tabular}{|cccccc|ccc|ccc|}
  \hline
\multicolumn{6}{|c|}{Contact network}&\multicolumn{3}{c|}{Estimate}&\multicolumn{3}{c|}{95\% C.I.} \\
 c1-c2 & c1-m & c1-d & c2-m& c2-d & m-d & MLE & pen.MLE & indep. &MLE & pen.MLE & indep. \\ 
  \hline
0 & 1 & 0 & 0 & 0 & 1 & 0.04 & 0 & 0 & [0, 0.11] & [0, 0] & [0, 0.01] \\ 
  0 & 1 & 1 & 0 & 0 & 1 & 0.04 & 0.06 & 0 & [0, 0.11] & [0, 0.14] & [0, 0.04] \\ 
  0 & 1 & 1 & 1 & 1 & 1 & 0 & 0.01 & 0.05 & [0, 0] & [0, 0.05] & [0, 0.26] \\ 
  1 & 1 & 1 & 0 & 0 & 1 & 0 & 0 & 0.02 & [0, 0] & [0, 0.04] & [0, 0.16] \\ 
  1 & 1 & 1 & 0 & 1 & 0 & 0 & 0.01 & 0.02 & [0, 0.15] & [0, 0.07] & [0, 0.23] \\ 
  1 & 1 & 1 & 0 & 1 & 1 & 0.07 & 0.08 & 0.13 & [0, 0.22] & [0, 0.20] & [0, 0.40] \\ 
  1 & 1 & 1 & 1 & 0 & 0 & 0 & 0.01 & 0.02 & [0, 0] & [0, 0.05] & [0, 0.21] \\ 
  1 & 1 & 1 & 1 & 0 & 1 & 0.05 & 0.06 & 0.1 & [0, 0.18] & [0, 0.16] & [0, 0.35] \\ 
  1 & 1 & 1 & 1 & 1 & 0 & 0.14 & 0.12 & 0.09 & [0, 0.45] & [0, 0.32] & [0, 0.52] \\ 
  1 & 1 & 1 & 1 & 1 & 1 & 0.65 & 0.65 & 0.54 & [0.30, 0.95] & [0.41, 0.89] & [0.05, 0.90] \\ 
   \hline 
\end{tabular}
\end{table}

\begin{table}
\caption{Estimated probability distribution of contact network for households with two 12--18 year olds and two 36+ year olds.  Dyad-independent, penalized likelihood (CV-selected $\lambda=199$), and unpenalized likelihood estimates are shown.}
\centering
\begin{tabular}{|cccccc|ccc|ccc|}
  \hline
\multicolumn{6}{|c|}{Contact network}&\multicolumn{3}{c|}{Estimate}&\multicolumn{3}{c|}{95\% C.I.} \\
 c1-c2 & c1-m & c1-d & c2-m& c2-d & m-d & MLE & pen.MLE & indep. &MLE & pen.MLE & indep. \\ 
  \hline
 0 & 0 & 0 & 1 & 1 & 0 & 0.07 & 0.01 & 0 & [0, 0.12] & [0, 0.07] & [0, 0.07] \\ 
  0 & 0 & 1 & 0 & 0 & 1 & 0.05 & 0 & 0 & [0, 0.14] & [0, 0.02] & [0, 0.02] \\ 
  0 & 0 & 1 & 1 & 1 & 1 & 0.03 & 0.04 & 0.03 & [0, 0.12] & [0, 0.12] & [0, 0.22] \\ 
  0 & 1 & 0 & 0 & 0 & 0 & 0.06 & 0 & 0 & [0, 0.11] & [0, 0] & [0, 0.01] \\ 
  0 & 1 & 1 & 1 & 1 & 1 & 0 & 0.03 & 0.06 & [0, 0.02] & [0, 0.11] & [0, 0.3] \\ 
  1 & 0 & 0 & 1 & 1 & 1 & 0 & 0.02 & 0.04 & [0, 0] & [0, 0.08] & [0, 0.27] \\ 
  1 & 0 & 1 & 0 & 0 & 1 & 0.07 & 0.01 & 0 & [0, 0.17] & [0, 0.09] & [0, 0.05] \\ 
  1 & 0 & 1 & 1 & 1 & 0 & 0.04 & 0.04 & 0.04 & [0, 0.21] & [0, 0.14] & [0, 0.29] \\ 
  1 & 0 & 1 & 1 & 1 & 1 & 0 & 0.12 & 0.11 & [0, 0.40] & [0, 0.28] & [0.01, 0.49] \\ 
  1 & 1 & 0 & 1 & 1 & 0 & 0.11 & 0.03 & 0.02 & [0, 0.23] & [0, 0.11] & [0, 0.23] \\ 
  1 & 1 & 0 & 1 & 1 & 1 & 0 & 0.07 & 0.07 & [0, 0.19] & [0, 0.19] & [0, 0.38] \\ 
  1 & 1 & 1 & 1 & 0 & 1 & 0 & 0.03 & 0.05 & [0, 0.11] & [0, 0.11] & [0, 0.26] \\ 
  1 & 1 & 1 & 1 & 1 & 0 & 0 & 0.07 & 0.07 & [0, 0.17] & [0, 0.19] & [0, 0.41] \\ 
  1 & 1 & 1 & 1 & 1 & 1 & 0.56 & 0.34 & 0.22 & [0, 0.70] & [0.14, 0.54] & [0.02, 0.67] \\ 
   \hline
\end{tabular}
\label{tab:hellindC}
\end{table}

\begin{figure}[tbp] 
  \caption{Estimated probability distribution for households with two 0--5 year olds and two 19+ year olds.  Labels are: ch1 = younger child, ch2 = older child, ad1 = female adult, ad2 = male adult}
  \centering
  \includegraphics[width=5.32in,height=3.32in,keepaspectratio]{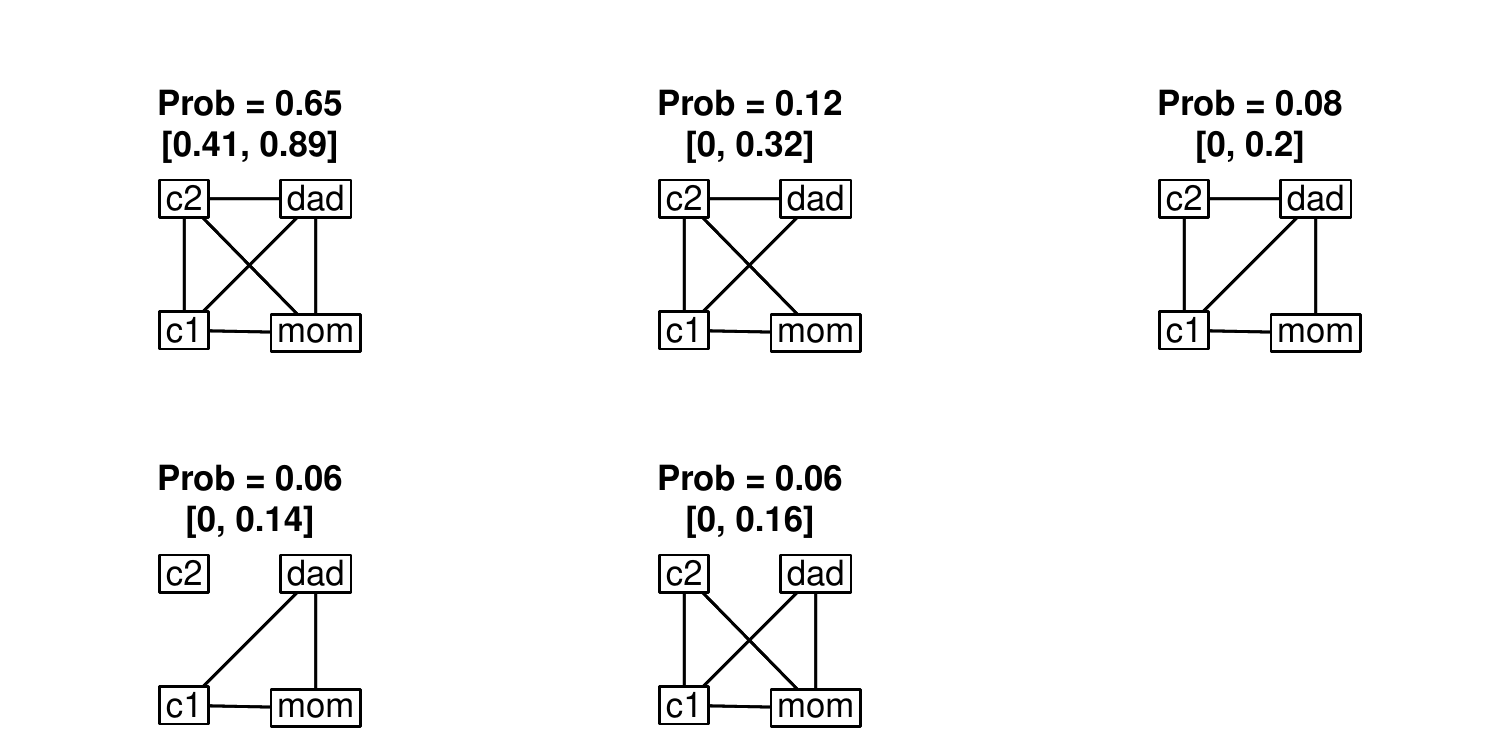}
  \label{fig:pen_hh_G}
\end{figure}

Figure~\ref{fig:pen_hh_G} displays the estimated probability distribution for contact networks in households with two 0--5 year olds and two 19+ year olds.  In other words, this figure graphically displays the penalized likelihood estimates in Table~\ref{tab:hellindG}.  Networks with estimated probabilities less than 0.03 are omitted from the plot.  The complete network has an estimated probability of 0.65.  The next most likely network has all contacts except contact between the two adults, and has an estimated probability of 0.12.  The third most likely network includes all contacts except between the elder child and the female adult, and has an estimated probability of 0.08.  The fourth most likely network has the elder child as an isolate, with all possible contacts occurring between the other three members.  Prior to analyzing this data, we would not have expected this network to have a non-negligible probability in households with such young children, as they require parental care.  However, it fits with two observations in our data set in which the elder 0--5 year old child was the respondent and reported no ties to other family members.  We hypothesize that the child was not at home on the survey date.  Since respondents were identified in advance of the survey date and mailed paper diaries to carry with them on the specified day, they were not necessarily at home.  This isolate effect is one source of dependency in our network estimates.  The plots show that in the five most likely networks, representing 97\% of the probability mass, the elder child contacts two or three of the other three members, or contacts none of them.  Networks in which the elder child contacts a single member are very unlikely.  If the elder child contacts at least one other household member, then he or she is more likely to contact the other two.  

The estimated probability distribution is reasonable for a household with this age composition, since small children require frequent contact with other family members.  The plot shows departure from the random mixing assumption.  Under random mixing, networks with the same number of total contacts would have the same probability, but we see that three networks with five contacts have estimated probabilities 0.12, 0.08, and 0.06.  Three other networks with five contacts are not displayed as their probability estimates are less than 0.03.  

\begin{figure}[tbp] 
  \caption{Estimated probability distribution for households with two 12--18 year olds and two 36+ year olds. Labels are: ch1 = younger child, ch2 = older child, ad1 = female adult, ad2 = male adult}
  \centering
  \includegraphics[width=5.37in,height=5.63in,keepaspectratio]{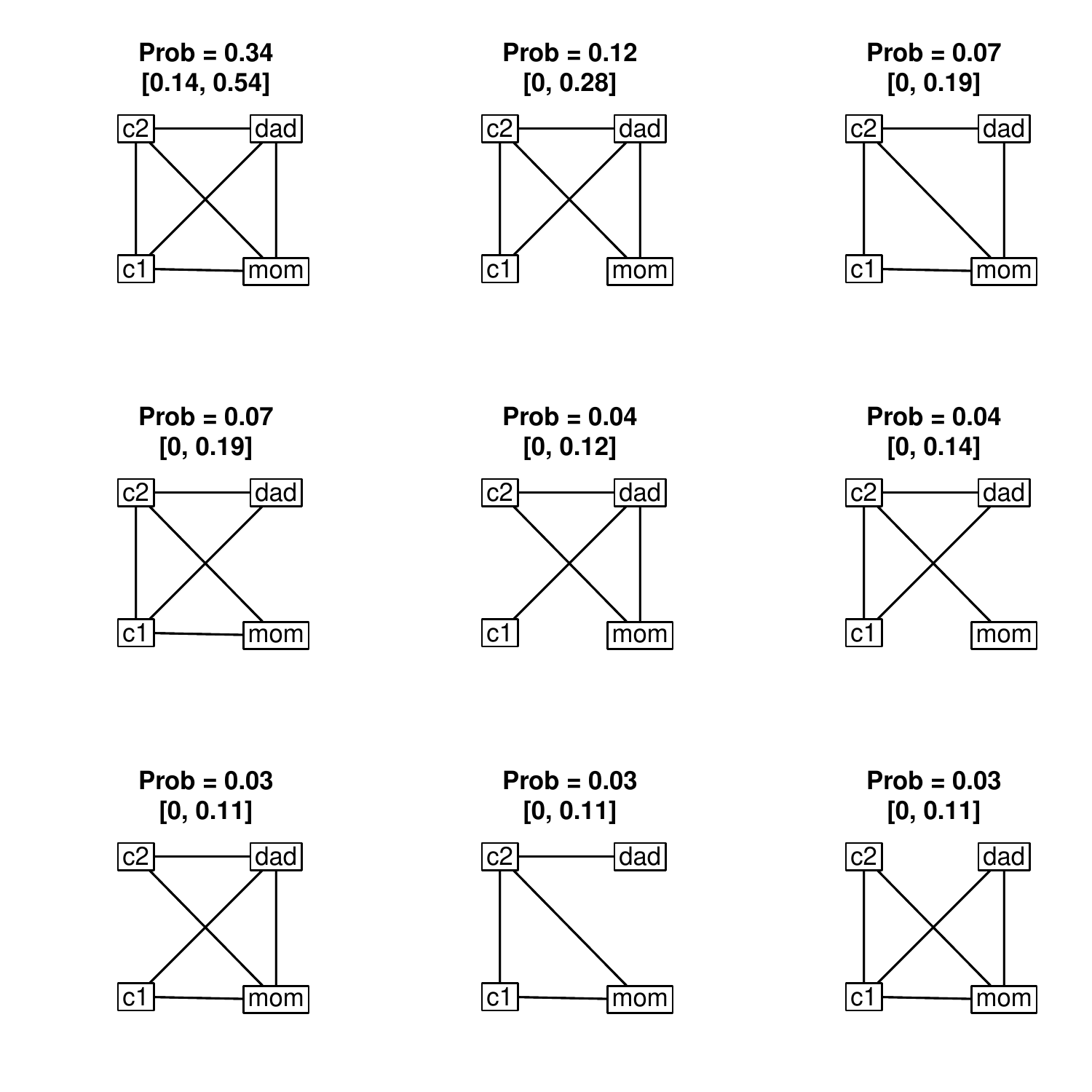}
  \label{fig:pen_hh_C}
\end{figure}

Figure~\ref{fig:pen_hh_C} shows the estimated probability distribution for households with two teenagers and two adults.  Here, the complete network is again the most likely network, but its estimated probability is substantially smaller: 0.34.  The second most likely network, whose probability estimate is 0.12, has all contacts except between the younger teenager and the female adult.  The third most likely network includes all contacts except between the younger child and the male adult.  This distribution also fits with our understanding of social behavior.  The teenage children are much less likely to contact their parents than 0--5 year old children.   By summing the probabilities of the complete network and the network in which all contacts occur except between the two adults, we see that there is only a 41\% chance that the two teenage children contact all other household members.  The corresponding probability is 77\% for the 0--5 year old children.

\begin{figure}[tbp] 
  \caption{Estimated probability distribution for households with two 19--35 year olds and two 36+ year olds. Labels are: ch1 = younger 19--35 year olds, ch2 = older 19--35 year old, ad1 = female 36+ year old, ad2 = male 36+ year old}
  \centering
  \includegraphics[width=5.38in,height=5.63in,keepaspectratio]{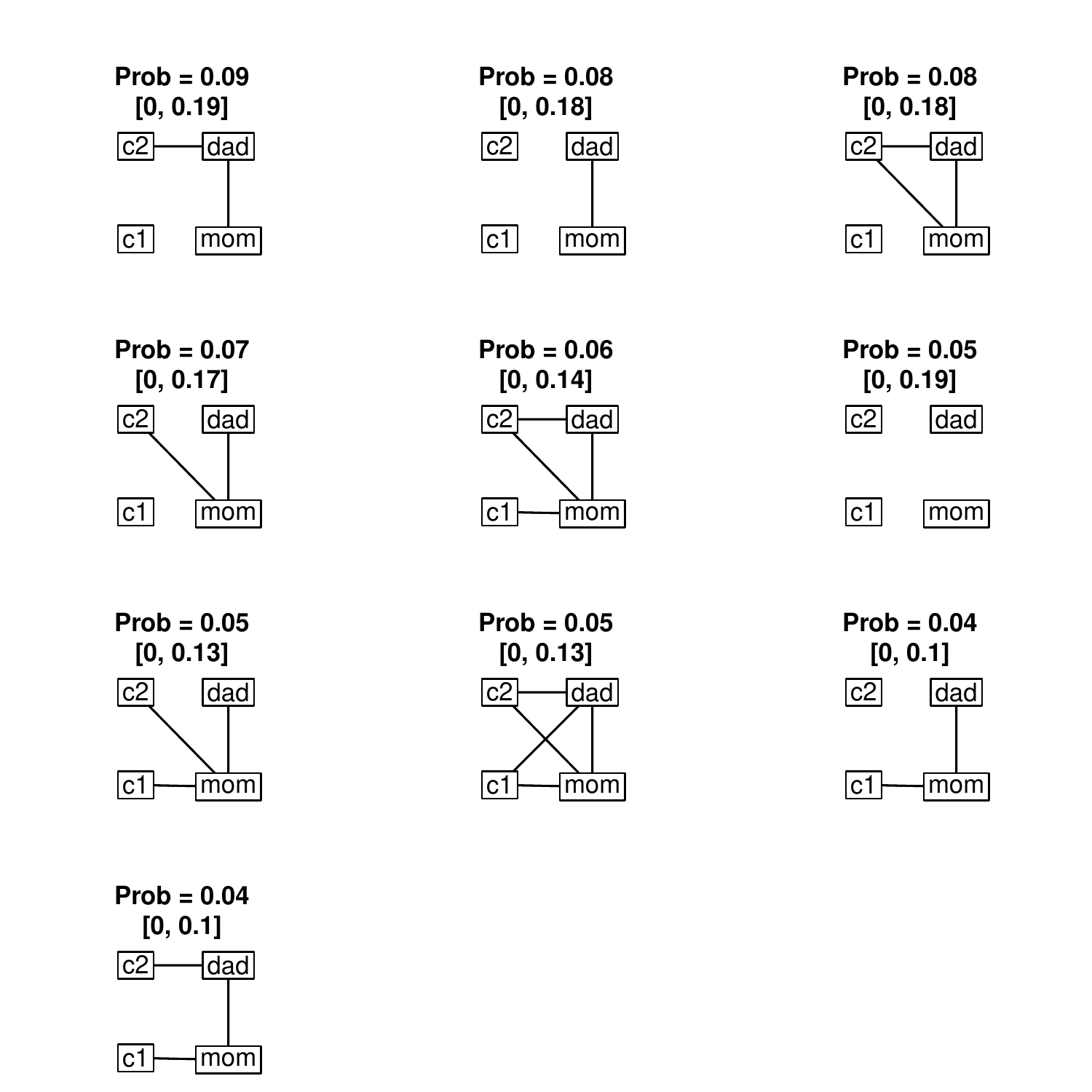}
  \label{fig:pen_hh_A}
\end{figure}

Figure~\ref{fig:pen_hh_A} shows the estimated probability distribution for households with two young adults and two older adults.  The network structure we observe is substantially different.  The most likely network is not the complete one, but a network with only two contacts, with probability 0.09.  The second most likely network includes only one contact: between the two older adults.  The empty network receives a non-negligible probability estimate of 0.05, unlike the younger households where the empty network received a negligible probability.  Furthermore, five of these eight most likely networks have one or more members as isolates.  The probabilities of these five networks sum to 41\%.  Since eight of 35 respondents reported no contacts to household members on the day of the survey, the high frequency of isolates in our final estimate is reasonable, and may also indicate that members were away from home.  Figures~\ref{fig:pen_hh_G}--~\ref{fig:pen_hh_A}  show a pattern of decrease in network density as the age of household members increases.
  
\subsection{Simulation study results}
\label{section:sim results}

\begin{figure}[!h]
\caption{Simulation results based on the characteristics of households with two 0--5 year olds and two 19+ year-olds.  The left-hand plot shows the mean squared error, squared bias, and variance averaged over probability parameters.  The right-hand plot shows the probability parameter estimates averaged over simulations.}\label{fig:simresults1}
\centering
\includegraphics[width=15cm,height=7.5cm]{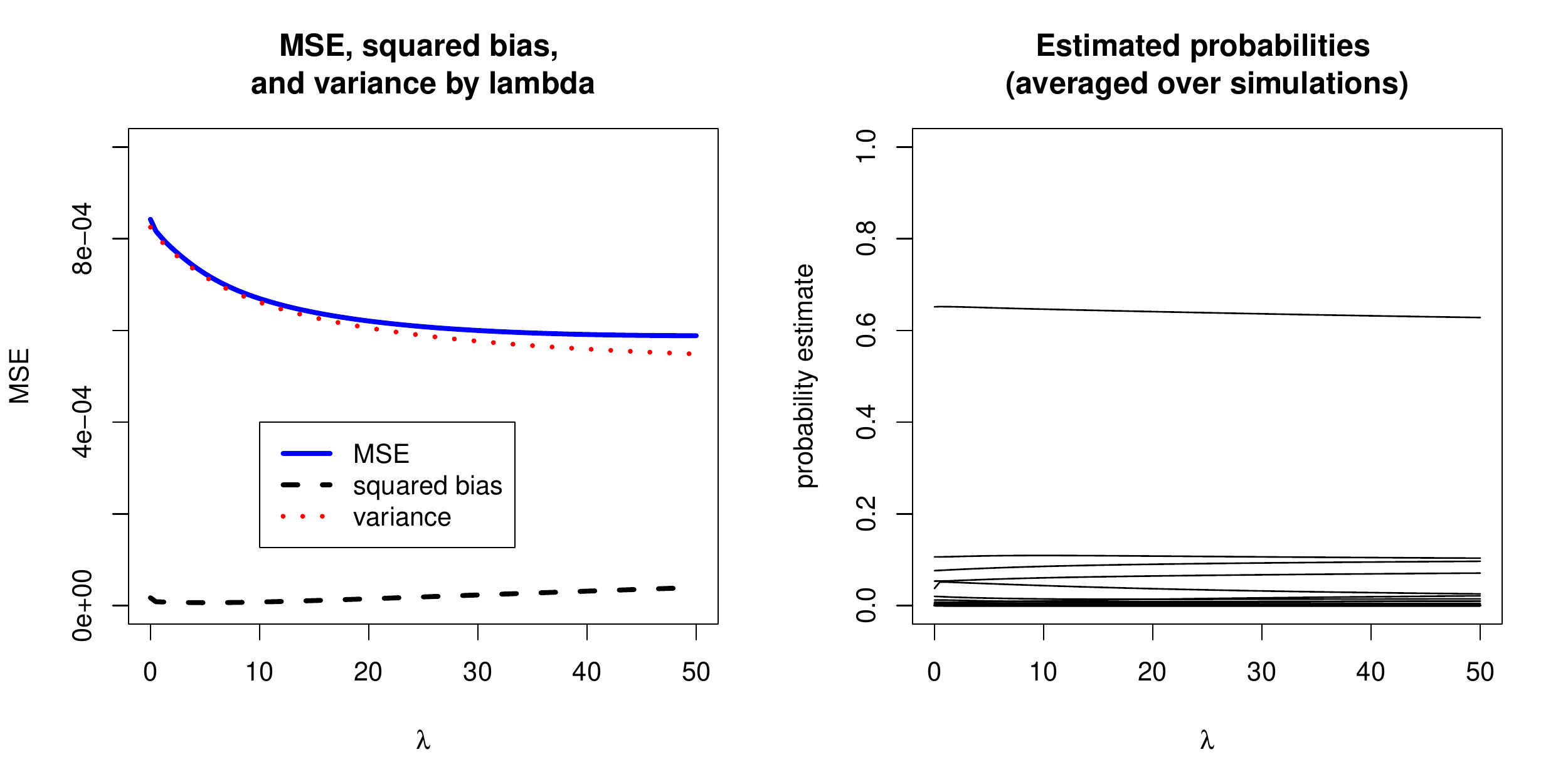}
\end{figure}

\begin{figure}[!h]
\caption{Simulation results based on the characteristics of households with two 12--18 year olds and two 36+ year-olds.  The left-hand plot shows the mean squared error, squared bias, and variance averaged over probability parameters.  The right-hand plot shows the probability parameter estimates averaged over simulations.}\label{fig:simresults2}
\centering
\includegraphics[width=15cm,height=7.5cm]{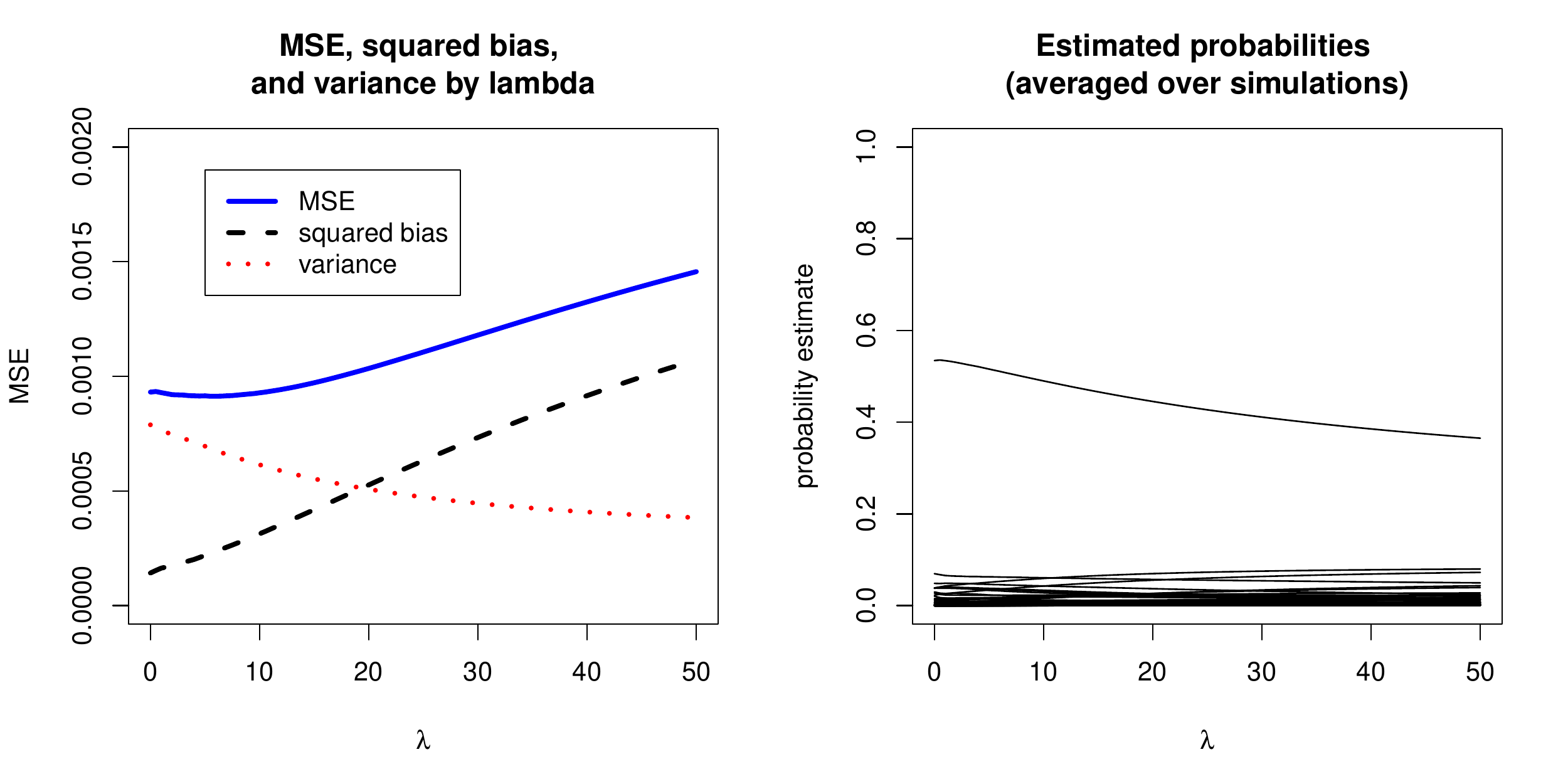}
\end{figure}

Figure~\ref{fig:simresults1} shows the mean average squared error and its bias-variance decomposition from simulations based on the characteristics of households with two 0--5 year olds and two 19+ year-olds.  After an initial decrease the mean averaged squared error stabilizes with increasing $\lambda$, due to decreasing bias and increasing variance.  The initial decrease in mean average squared error shows the improvement in predictive performance as the weight on the penalty term is increased.  The eventual stabilization of MSE shows similar predictive performance for a range of $\lambda$-values.  Figure~\ref{fig:simresults2} shows a different pattern for simulations based on the characteristics of households with two 12--18 and two 36+ year olds.  For this household composition, MSE shows a very small decrease and then increases.  The squared bias increases steadily while variance decreases monotonically.  The right-hand plots show that as $\lambda$ increases, the probability parameter estimates converge to the independence model estimates as we would expect.  

\clearpage
\section{Discussion}
\label{section:discussion}

We have used egocentric data to estimate within-household contact networks, a key component of epidemic spread.   We analyzed several different household types and found substantial differences in contact behavior between households of different age compositions.  Contact density decreased as members' age increased, suggesting that the higher transmission probabilities estimated for children than adults may be due to differences in contact behavior rather than biological differences.  We also found evidence for departure from the ``random mixing'' assumption commonly used in epidemic models.  The independence assumption required for random mixing does not hold, as our final estimates are different from the independence model estimates.  One possible source of contact dependency is an isolate effect, in which members who are not at home make no contacts to at-home household members.  One strength of our method is that it uses as few parametric assumptions as possible.  As such, our results can be used to build a parametric model based on the patterns we found or to assess assumptions made by existing models.  This work also contributes to the field of social network inference.  Using egocentric data collected from multiple small networks, we develop methodology to infer the probability distribution of the complete network with minimal assumptions.  Our method could be applied to network data with the same structure from other settings.  

Our method does require some assumptions.  Our choice of smoothing penalty imposes a preference for probability distributions that are similar to an independence model.  This is a lighter constraint than assuming independence, and permits dependence in our final estimates.  We found this penalty to work better than the other two we tried.  The adjacency penalty oversmoothed and produced unrealistic estimates, and the exchangeability penalty did not sufficiently constrain the parameter space.  

An alternate solution to the identifiability problem would be a Bayesian approach, in which we restrict the parameter space by expressing our beliefs about the parameter values through prior distributions.  However, the state of prior knowledge in the field is weak.  The only paper we know of inferring household contact networks is~\cite{potter}, and that paper uses the same data set we analyze here, so does not truly give prior knowledge.  Therefore, we prefer the penalized likelihood approach scientifically.  However, we did perform Bayesian analysis as an exploration.  A noninformative prior would not resolve the identifiability problem.  Our understanding of social behavior might motivate us to create a prior distribution imposing a preference for denser networks, since we expect most household members to contact each other on a given day.  However, Figure~\ref{fig:data2} shows that this prior distribution would be inappropriate for some household types.  We are estimating 63 dependent parameters, and a prior distribution placing large weight on denser networks necessarily places negligible weight on networks with zero, one, or two contacts.  Furthermore, the variance of the prior distribution for each parameter needs to be small, because priors with large variance were insufficient to constrain the parameter space.  Thus, networks which are actually fairly likely given this data set received negligible probability mass in the posterior.  We feel a Bayesian approach would be appropriate if we had a high level of confidence in our prior beliefs, and the exploration described here shows that the belief in denser household networks was not borne out by the data.  Our penalized likelihood approach succeeds because the penalty itself is informed by the data, so the constraint it imposes on the parameter space is compatible with our data.  


Our work has a number of limitations.  First, we made assumptions regarding which contacted individuals are household members since this information was not collected, and we made assumptions about the identity of each contacted person based on their reported age and sex.  In future surveys, we recommend having respondents identify which of their contacts are to household members.  In addition, since we found evidence that some household members are away from home on the survey date, we recommend collection of home/away status for each household member.  

Our approach is for networks of a fixed size and age composition and requires adequate sample size.  Our data set contains 750 respondents, but because we performed analyses separately for each age composition, sample sizes ranged from 23--40.  In two of the six household types we analyzed, the optimal tuning parameter was large and estimates were close to those assuming independence.  The high contribution of the penalty to the estimates indicates a high level of non-identifiability for these household types.  Our method works only for small networks because the proportion of the network observed from one respondent per household decreases quickly with network size.  In future surveys we recommend collecting contact reports from all household members to obtain the fullest possible understanding of the contact network.  Our nonparametric approach will directly apply to completely observed household networks, and without missing data, the penalty term will be unnecessary and inference will be straightforward.  In cases where nonresponse results in a small amount of missing data, the parameters may be identifiable with the nonparametric method.  If not, our penalized likelihood approach can be easily modified to accomodate reports from multiple respondents per household.  

In our analysis, we assumed that contact behavior is the same on weekdays and weekends, and during the Easter holiday versus a non-holiday period.  In fact, contact patterns may change during these periods, but sample sizes were too small to perform separate estimates since we performed estimates separately for each household age composition.  A parametric model based on explicit assumptions of contact behavior could use the entire data set to estimate patterns, thus increasing our power to detect weekend and holiday effects.  

One example of a parametric model was implemented in~\cite{potter}.  In that paper, the authors estimated a latent variable indicating whether each household member is at home on a given day.  They assumed the home/away statuses of the different members were independent, and that contacts occurred independently between members at home, with contact probabilities depending only on age.  They assumed that members away from home were not contacted.  One advantage of this approach is that they combined reports from households of different sizes and age compositions, so increasing the sample size, while estimating a smaller number (20) of parameters.  By estimating fewer parameters with a larger sample size, they were also able to estimate separate network effects for weekday vs. weekend and holiday vs. non-holiday.  They found no evidence for differences in contact patterns between the weekday and the weekend.  They found that holiday and non-holiday parameter estimates were statistically different, but did not show a clear and substantively important pattern in the differences.  The disadvantage to the parametric model is the large number of assumptions required.  In this paper, our goal was to perform estimation with as few assumptions as possible.  The approach outlined here is well suited to that purpose, and is preferable when we have limited prior knowledge about our parameters of interest and a large amount of data.  We recommend the parametric approach when researchers feel confident that model assumptions hold.

We have developed a new technique to infer small contact networks from egocentric data using minimal assumptions and applied it to estimate household contact networks in Belgium.  Our estimates show departure from the random mixing assumption found in many epidemic models.  We recommend collecting additional contact data and further investigation of the contact network structure and its relevance for infectious disease transmission.

\subsection{Acknowledgements}
We are grateful to Mark S. Handcock, Ira M. Longini, Jr. and M. Elizabeth Halloran for providing their comments on this research.  We thank the POLYMOD project for providing the data we analyzed.  We thank the NIH/NIGMS MIDAS grant U01-GM070749 for funding this research.  For the simulations we used the infrastructure of the VSC - Flemish Supercomputer Center, funded by the Hercules foundation and the Flemish Government - department EWI.

\bibliographystyle{chicago}
\bibliography{thesis}

\end{document}